\documentclass[12pt]{article}
\usepackage{graphicx, amsmath, amssymb, cite, setspace, color, relsize, bm, bbold, array, hyperref, dsfont, bbold, xcolor, cancel,soul}

\usepackage[top=1 in, bottom=1 in, left=.9 in, right=.9 in]{geometry}

\newcommand{\captionfonts}{\footnotesize} 
\makeatletter
\long\def\@makecaption#1#2{%
  \vskip\abovecaptionskip
  \sbox\@tempboxa{{\captionfonts #1: #2}}%
  \ifdim \wd\@tempboxa >\hsize
    {\captionfonts #1: #2\par}
  \else
    \hbox to\hsize{\hfil\box\@tempboxa\hfil}%
  \fi
  \vskip\belowcaptionskip}
\makeatother

\def\lsim{ \lower .75ex \hbox{$\sim$} \llap{\raise .27ex
\hbox{$<$}} }
\def\gsim{ \lower .75ex \hbox{$\sim$} \llap{\raise .27ex
\hbox{$>$}} } 
\def\a{h}

\let\oldsqrt\sqrt
\def\sqrt{\mathpalette\DHLhksqrt}
\def\DHLhksqrt#1#2{%
\setbox0=\hbox{$#1\oldsqrt{#2\,}$}\dimen0=\ht0
\advance\dimen0-0.2\ht0
\setbox2=\hbox{\vrule height\ht0 depth -\dimen0}%
{\box0\lower0.4pt\box2}}

\begin{document}

\title{Hyperinflation}

\author{Adam~R.~Brown  \\
 \textit{\small{Physics Department, Stanford University, Stanford, CA 94305, USA}} }
\date{}
\maketitle
 \let\thefootnote\relax\footnotetext{Published in PRL {\bf 121}, 251601 as ``Hyperbolic Inflation''.}

\vspace{-.95cm}


\begin{abstract}
\noindent A  model of cosmological inflation is proposed in which field space is a hyperbolic plane. The inflaton never slow-rolls, and instead orbits the bottom of the potential, buoyed by a centrifugal force.  
Though initial velocities  redshift away during inflation, in negatively curved spaces angular momentum naturally starts exponentially large and remains relevant throughout. 
Quantum fluctuations produce perturbations that are adiabatic and approximately scale invariant; strikingly, in a certain parameter regime the perturbations can grow double-exponentially during horizon crossing. 
\end{abstract}

\thispagestyle{empty} 
\newpage

Negative curvature makes geodesics diverge. When aiming for a distant target, this makes it easy to miss: the required accuracy grows exponentially with distance. In this paper, the `target' being missed will be the bottom of the inflaton potential. By giving  field space the geometry of the hyperbolic plane, the inability of the inflaton to easily find its minimum will prolong inflation long enough to flatten the universe.

In orthodox slow-roll inflation \cite{slowroll}, field space is a line $\phi(\vec{x},t) \in \mathbb{R}^1$ and the field has action
\begin{equation}
S_{\mathbb{R}^1} = \int dt \, d^3 x \, a(t)^3 \left(  - \frac{1}{2}  (\nabla_{\mu} \phi )^2  -V(\phi) \right) , \label{eq:SR1}
\end{equation}
where $a(t)$ is the FRW scale factor $ds^2 = -dt^2 + a(t)^2 d\vec{x}^2$. The expansion of the universe introduces `Hubble friction' into the equation of motion for the homogeneous  mode of the field $\ddot{\phi} + 3 H \dot{\phi} = -V'(\phi)$, which has the effect of 
 redshifting away 
 field velocity
and thus slowing the inflaton's slide to the bottom of 
the potential. Nevertheless, for inflation to last the required sixty $e$-folds, $V(\phi)$ must be tuned to be flat.

The  potential would not need to be so flat were Hubble friction assisted by another retarding force. 
For example, a centrifugal force may arise when field space is two-dimensional
\begin{equation}
S_{\mathbb{R}^2} = \int dt  \, d^3 x \,  a(t)^3 \left( - \frac{1}{2}    (\nabla_{\mu} \phi )^2 -  \frac{1}{2} \phi^2    (\nabla_{\mu} \psi )^2 -V(\phi) \right) .
\end{equation}
Here $\phi$ is the `radial' field direction, $\psi$ is the `angular' field direction, and the rotationally symmetric potential has a minimum at $\phi =0$. 
Rather than rolling straight down the potential, the field now orbits the bottom, and inflation cannot end until the inflaton has lost its angular momentum.  This prolongs inflation, but not by much: Hubble friction makes all velocities exponentially redshift away, so after a few $e$-folds angular motion is irrelevant \cite{Easson:2007dh}. 

In this paper, I will show that angular motion can remain relevant throughout inflation if field space is a hyperbolic plane (with curvature length $L$)
\begin{equation}
S_{\mathbb{H}^2} = \int dt  \, d^3 x \,  a(t)^3 \left( -   \frac{1}{2}  (\nabla_{\mu} \phi)^2 - \frac{1}{2} L^2 \sinh^2 \frac{\phi}{L}   (\nabla_{\mu} \psi)^2  -V(\phi) \right) . \label{eq:actionH2}
\end{equation}
The negative curvature plays an essential role.
 In $\mathbb{R}^2$, to have exponentially large field-space angular momentum $\phi^2 \dot{\psi}$ requires either the distance $\phi$ or the orbital kinetic energy density $\rho = \frac{1}{2} \phi^2 \dot{\psi}^2$ to be exponentially large. In a negatively curved space this is no longer true. In $\mathbb{H}^2$, the field-space angular momentum  
\begin{equation}
 J  =  L^2 \sinh^2 \frac{\phi}{L}  \, \dot{\psi}
\end{equation}
may be exponentially large even while both $\phi$ and $\rho = \frac{1}{2} L^2 \sinh^2 \frac{\phi}{L}  \, \dot{\psi}^2$ stay moderate. Since the angular momentum starts off exponentially large it can remain relevant even as it redshifts.

The equation of motion for the homogeneous mode of $\psi$ gives a conserved quantity $ J (0) = a(t)^3  J (t)$. 
The equation of motion for the homogeneous mode of $\phi$ is 
\begin{equation}
\ddot{\phi} + 3 H \dot{\phi} - L \sinh \frac{\phi}{L} \cosh \frac{\phi}{L} \dot{\psi}^2 = - \partial_\phi V(\phi) \label{eq:eomforphi} .
\end{equation}
Let us look at the attractor solution that asserts itself after a few Hubble times when $V$ is slowly varying (so that we may neglect the $\ddot{\phi}$ term) 
and $\phi \gg L$. When the slope is shallow, $\frac{\partial_\phi V}{3H} < 3 H L$,  the radial field velocity $\dot{\phi}$ is controlled by Hubble friction
\begin{equation}
\textrm{slow-roll attractor:} \ \ \ \ \ \dot{\phi} = - \frac{\partial_\phi V }{3H}; \ \ \ \ \ \dot{\psi} = 0 .   \ \ \ \ \ \ \ \ \ \ \ \  \ \ \ \ \ \ \  \ \  \ \ \ \ \ \ \  \label{srattractor}
\end{equation}
This is slow-roll inflation, as in  Eq.~\ref{eq:SR1}. When the slope is steep, $\frac{\partial_\phi V}{3H} > 3 H L$, 
the radial field velocity is controlled by the centrifugal force 
\begin{equation}
\textrm{hyperinflation attractor:} \ \ \ \ \  \dot{\phi} =  -3 H L ; \ \  \ \ \ L \sinh \frac{\phi}{L} \dot{\psi} =   \sqrt{L \, \partial _\phi V - (3 H L)^2}  . \label{hiattractor}
\end{equation}
In the hyperinflation attractor, the rate at which $\phi$ approaches the origin is independent of the slope, and is set only by how fast the field can dump angular momentum. 

For $\frac{\partial_\phi V}{3H} > 3 H L$, slow-roll inflation is unstable, since any angular motion---even that sourced by quantum fluctuations---gets exponentially magnified. The field will soon find itself in the hyperinflation attractor and stay there for a number of $e$-folds equal to 
\begin{equation}
\# \textrm{$e$-folds} = \frac{\phi_\textrm{initial}}{3L} . \label{eq:numberofeolds}
\end{equation}
So long as $L \partial_{\phi} V \ll V$, the kinetic energy density in the homogeneous motion of the field,
\begin{equation}
\rho = \frac{1}{2} \dot{\phi}^2 + \frac{1}{2} L^2 \sinh^2 \frac{\phi}{L} \dot{\psi}^2 = \frac{1}{2} L \, \partial_\phi V,
\end{equation}
is much less than the potential energy, and the Hubble constant, $H^2 \sim G V/3$, varies only slowly. \\
%
Hyperinflation can thus provide dozens of $e$-folds of exponential expansion, solving the zeroth-order problems that first motivated Guth \cite{Guth:1980zm}. \\

Hyperinflation can also solve the first-order problem of seeding structure. 
Perturbations in the logarithm of the angular momentum are conserved outside the horizon, and since it is precisely the angular momentum that is delaying the end of inflation, these perturbations are adiabatic.  As shown in  the appendix, the resultant density fluctuations have
\begin{equation}
\frac{\delta \rho}{\rho} \sim \frac{H^2}{\dot{\phi}} e^\a \Bigl|_\textrm{horizon exit} \sim \frac{H}{L} e^\a \Bigl|_\textrm{horizon exit},  \ \ \ \  \textrm{where } \a \equiv 3 \sqrt{\frac{\partial_\phi V}{9H^2L}-1}. \label{eq:perturbationsummary}
\end{equation}
For sufficiently slowly varying $h$ and $H$, the perturbations are scale invariant. 

The factor of $e^\a$ is remarkable, and I will remark on it now. 
As we will see in the appendix,  modes are well behaved deep inside the horizon, and well behaved outside the horizon, but grow like $e^{e^{Ht}}$ during the $\log [\sqrt{2} \a]$ $e$-folds immediately preceding horizon exit.
This double-exponential growth arises because angular momentum and angular velocity are \emph{anti}correlated. At fixed radius more angular momentum would mean more angular velocity, but the radius is not fixed. Instead, more angular momentum means more centrifugal force, which means larger radius, which in turn means less angular velocity. (This effect is familiar from orbital dynamics: if the International Space Station fires a rocket to increase its angular momentum, it gains altitude and its orbital period lengthens.) Regions with larger angular velocity drag along regions with smaller angular velocity, so gradient couplings transfer angular momentum from patches with less to patches with more, leading to explosive growth.

Tensor perturbations are not magnified (they're still given by the standard formula  $\Delta^2_t \sim 2 V/3 \pi^2 M_\textrm{Pl}^4$ \cite{Baumann:2009ds}) and so for $\a \, \gsim  \,5$ the tensor-to-scalar ratio is unobservably tiny\footnote{Indeed this is why, insufficiently courageous in my convictions, I shelved this paper when the BICEP result was announced.}. \\

Let us now examine the radiative stability of the action, Eq.~\ref{eq:actionH2}. Superficially it looks like, even ignoring the potential, there is already a problem with the kinetic term. Taylor expanding $\sinh \frac{\phi}{L}$, the principal contribution near $\phi = 3N L$ comes from the $\phi^{3N}$ term, which from this perspective is a very non-renormalizable operator. However, what this naive power-counting cannot see is that the geometry of field-space is maximally symmetric, and that the three hyperbolic symmetries of $\mathbb{H}^2$ are of equal dignity with those of $\mathbb{R}^2$ and protect the form of the kinetic term against quantum corrections. The only correction against which this symmetry cannot protect is the only correction that does {not} break the symmetry: renormalization of $L$. 

Since $L$ is the scale in the denominator of an irrelevant operator, it is natural to expect it to be of order the UV cutoff of the effective field theory. Quantum gravity tells us that this UV cutoff should be no \emph{larger} than the Planck scale, but there is no reason the cutoff cannot be \emph{smaller}, which is good because for small-field hyperinflation we desire $L \, \lsim \, 10^{-3} \, M_\textrm{Pl}$. (This situation is directly analogous to that of the axion decay constant, which is another energy scale that appears in the denominator of an irrelevant operator: while it is difficult for the decay constant to be super-Planckian \cite{Banks:2003sx}, there is no difficulty for it to be  sub-Planckian.) 
Indeed, a similar situation arises in the model of compact hyperbolic extra dimensions considered in \cite{Kaloper:2000jb}, where, as in my model, the curvature scale (in units of the brane tension) is meant to be much shorter than the four-dimensional Planck scale $M_\textrm{Pl}$. In \cite{Kaloper:2000jb} this is achieved by the intercession of a third scale, $M_*$, the fundamental scale of the higher-dimensional theory, which provides a natural cutoff on the running of $L$. The hyperbolic geometry of the extra dimensions means that even at moderate linear size the extra dimensions have exponentially large volume, which helps stabilize $M_* \ll M_\textrm{Pl}$ and so $L \ll M_\textrm{Pl}$. It would be interesting to embed hyperinflation in this model and further explore the radiative stability of the kinetic term  (see also \cite{Mizuno:2017idt}).

The potential term breaks the hyperbolic symmetry, and so can induce even radiative corrections that do not respect the hyperbolic symmetry. As a foil,  consider slow-roll inflation in a potential
\begin{equation}
V(\phi) = \alpha^2 M_\textrm{Pl}^4 \left( 1 + \beta \frac{\phi}{M_\textrm{Pl}} + \frac{1}{2} \eta \left( \frac{\phi}{M_{\textrm{Pl}}} \right)^2 + \ldots  \right) . \label{eq:potentialexpansion}
\end{equation}
In small-field slow-roll inflation, $\alpha$ is tiny, let's say $10^{-20}$. For modes that cross the horizon near $\phi =0$, the scalar perturbation amplitude is $\frac{\delta \rho}{\rho} \sim \frac{H^2}{\dot{\phi}} \sim \frac{\alpha^2 M_\textrm{Pl}^2}{\alpha \beta M_\textrm{Pl}^2} \sim 10^{-5}$, so the potential must be shallow $\beta \sim  10^{5} \alpha$. More troublingly, not only must the potential be shallow, it must remain shallow for many Hubble times: each $e$-fold, the field advances by $\Delta \phi = H^{-1} \dot{\phi} \sim \beta M_\textrm{Pl}$, which is observationally constrained to change the perturbation amplitude by only a few percent ($n_\textrm{s} \sim 0.97$), so $\eta \frac{\Delta \phi}{M _\textrm{Pl}} \sim 10^{-2} \beta$  and therefore $\eta \sim 10^{-2}$. That $\eta$ needs to be smaller than its `natural' value of $1$ is known as the `$\eta$ problem'. 


Now let's consider hyperinflation in the same potential. In hyperinflation, $\beta$ does not need to be nearly so small, since even if the perturbations start off unacceptably tiny, they can be exponentially magnified during horizon crossing.
However, the same exponential magnification makes the size of the perturbations very sensitive to changes in the slope of the potential. In order to get only a gentle percent-level running of the magnitude of perturbations, 
we need $\a$ to change by no more than $10^{-2}$ per $e$-fold. Since during a single $e$-fold $\Delta \phi = 3L$, this means that $\eta \sim 10^{-2}$. Unlike in slow-roll inflation, the potential does not need to be shallow to get suitably sized perturbations; but like in slow-roll inflation, approximate scale invariance means the potential does need an unnaturally small second-derivative.

Quite how much of a problem this is is hard to say. A symmetry-breaking $V(\phi)$ will induce loop corrections both to itself and to the kinetic term. However, radiative corrections arise from large field deviations, and are therefore sensitive to the curvature of field-space, so the standard $\mathbb{R}^2$ analysis is unreliable. Since this is a question of radiative stability in the face of Planck-scale corrections, it can only definitively be addressed in the context of a UV-complete theory of quantum gravity.  Hyperbolic field spaces are known to arise in such theories \cite{Kaloper:2000jb,Horne:1994mi} and it would be interesting to try to embed hyperinflation in such a setting.

Finally, notice that even though the action Eq.~\ref{eq:actionH2} makes reference to an exponentially large field range, the range actually expressed 
over the course of inflation is safely sub-Planckian \cite{Kallosh:1995hi}.\\

The exponential dependence on $\a$ means that small changes in 
the potential can give rise to large changes in the perturbation magnitude. As we have seen, this can make the flatness of the power spectrum delicate; but also this bug can become a primordial feature---for example, a late burst of large-$\a$ hyperinflation could dump energy into small black holes \cite{GarciaBellido:1996qt}.

Another natural place to look for a signature of hyperinflation would be  higher-point perturbation statistics. The exponential magnification of the power spectrum does not in-of-itself produce non-Gaussianities since multiplication (even by an exponentially large number) is linear. Indeed, 
 the squeezed limit of the bispectrum should be small and satisfy the single-field consistency condition
\cite{Maldacena:2002vr,Creminelli:2004yq}
\begin{equation}
f_\textrm{NL} = \frac{5(n_\textrm{s} -1)}{12} .
\end{equation}
However, the other bispectrum shapes are not so constrained and it would be interesting to investigate, for example, the equilateral non-Gaussianity (using techniques drawn from \cite{Chen:2009we,Chen:2009zp,Baumann:2011su,Achucarro:2010jv,Achucarro:2010da,Gong:2011uw,Kaiser:2012ak,Assassi:2013gxa,Kidani:2014pka,Tolley:2009fg}). \\

Non-canonical kinetic terms are by now a familiar part of inflationary model-building, and ideas with some of the same ingredients as hyperinflation include \cite{Easson:2007dh,Tolley:2009fg,Gordon:2000hv,GrootNibbelink:2001qt,Berg:2009tg,Adshead:2012kp,Jenkins:2012tq,GarciaBellido:2011de,Adshead:2016iix} and particularly \cite{Cremonini:2010ua}. A prominent example of a hyperbolic inflaton field space is the ``$\alpha$-attractor'' scenario of \cite{Carrasco:2015uma}, but for that model the potential is so flat that it is firmly in the slow-roll regime of Eq.~\ref{srattractor}. \\

Hyperinflation can both flatten the universe (Eq.~\ref{eq:numberofeolds}) and seed its structure (Eq.~\ref{eq:perturbationsummary}). In the appropriate parameter regime it's not optional, it's compulsory---when hyperinflation is an attractor, slow roll is a repeller (see also \cite{Renaux-Petel:2015mga}). Two natural next steps have been discussed: investigating the radiative stability against Planck corrections; 
and exploring potential observational signatures from higher-point primordial statistics. Hyperinflation is dramatically different from slow-roll inflation: the kinetic motion is now always relevant; the field now never slow rolls; and perhaps most strikingly, during horizon exit, as the scale factor exponentially inflates, the perturbations double-exponentially hyperinflate. 

\section*{Acknowledgements}
Thanks to Daniel Baumann, Robert Brandenberger, Daniel Green, Shamit Kachru, Renata Kallosh, Andrew Liddle, Andrei Linde, David Marsh, Mehrdad Mirbabayi, Subodh Patil, Diederik Roest, Eva Silverstein, an anonymous referee, and the hosts of PrimoCosmo13 at KITP. 

\appendix

\section{Perturbations}
For a sufficiently slowly varying potential, we may treat the background as  fixed de Sitter space
\begin{equation}
ds^2 = -dt^2 + e^{2Ht} d\vec{z}^2  =  \frac{ -d \tau^2 +  d\vec{z}^2 }{H^2 \tau^2}  , \label{conformal} 
\end{equation}
where $H \tau \equiv - e^{-H t}$, and treat the slope $\partial_\phi V$ as fixed. Then for $\phi \gg L$ the action Eq.~\ref{eq:actionH2} becomes
\begin{equation}
S = \int dt \, d^3z \, e^{3Ht}  \left( \frac{1}{2} \dot{\phi}^2   - e^{-2Ht} \frac{1}{2} ( \partial_{\vec{z}}  \phi)^2  + L^2 e^{2 \phi/L}  \left( \frac{1}{2} \dot{\psi}^2   - e^{-2Ht} \frac{1}{2}  ( \partial_{\vec{z}} \psi)^2 \right) - (\partial_\phi V) \phi \right).
\end{equation}
For $\partial_\phi V> 9 H^2 L$, this gives rise to the hyperinflation attractor\footnote{As explained around Eq.~\ref{hiattractor}, the hyperinflation path is very different from the slow-roll path: rather than slumping to the bottom via gradient descent, the field now spirals in on a decaying orbit. But note that, \emph{constrained to the hyperinflation path}, a slow-roll relation does link the field velocity  to the rate of change of $V$.}, Eq.~\ref{hiattractor}
\begin{eqnarray}
\phi &=& - 3 H L t + e^{-Ht} \Phi(\vec{x},t) \\
 \psi &=& \frac{\sqrt{L \partial_\phi V  - {9 H^2 } L^2}}{3 H L  } e^{3Ht}  +  L^{-1} e^{2Ht}   \Psi(\vec{x},t) .
\end{eqnarray}
Fourier transforming $\Phi(\vec{x},t) = \int d \vec{k} \, e^{i \vec{k} \cdot \vec{x}}  \Phi_{\vec{k}} (t)$ and then expanding the action to first order in the perturbations gives a total derivative (as it must, since we are perturbing around a classical solution); expanding to second order gives 
\begin{equation}
S =  \sum_{\vec{k}} \int d \tau \left[  \frac{1}{2} \dot{\Phi}^2_k  + \frac{1}{2} \dot{\Psi}_k^2  - \frac{2\a}{\tau} \Phi_k \dot{\Psi}_k + \frac{1 + \a^2}{\tau^2} \Phi_k^2 + \frac{4 \a}{\tau^2} \Phi_k \Psi_k + \frac{1}{\tau^2} \Psi_k^2    - \frac{1}{2} k^2 \Phi_k^2 - \frac{1}{2} k^2 \Psi_k^2  \right] , \nonumber
\end{equation}
where dots differentiate with respect to $\tau$ and $\a \equiv 3 \sqrt{\frac{\partial_\phi V}{9H^2L}-1}$. 
%
%
The equations of motion are 
\begin{eqnarray}
\ddot{\Phi}_k + \frac{2 \a}{\tau} \dot{\Psi}_k - \frac{4 \a}{\tau^2} \Psi_k - \frac{2\a^2}{\tau^2} \Phi_k - \frac{2}{\tau^2} \Phi_k + k^2 \Phi_k & =&  0 \label{eq:xeom1} \\
\ddot{\Psi}_k - \frac{2 \a}{\tau} \dot{\Phi}_k - \frac{2}{\tau^2} {\Psi}_k - \frac{2\a}{\tau^2} \Phi_k + k^2 \Psi_k & = & 0. \label{eq:yeom1}
\end{eqnarray} 
[Equation~\ref{eq:yeom1} may also be expressed in terms of the Fourier modes of the  angular momentum, 
\begin{equation}
\frac{d (a^3  J _k)}{d \tau} = - H^2 k^2  \tau^2 \Psi_k  \label{eq:reteofchangeofGammaperturbation} ,
\end{equation}
confirming that $a^3  J$ is conserved far outside the horizon ($k$=0).]

Perturbations well inside the horizon ($|k \tau| \gg 1$) are given by
\begin{eqnarray}
\Psi_k & = & C_1 e^{i k \tau + i \a \log k \tau} + C_2 e^{i k \tau - i \a \log k \tau} + C_3 e^{-i k \tau + i \a \log k \tau} + C_4 e^{-i k \tau - i \a \log k \tau}  \label{early1}  \\
\Phi_k & = & C_1 i  e^{i k \tau + i \a \log k \tau} + C_2 (-i) e^{i k \tau - i \a \log k \tau} + C_3 i e^{-i k \tau + i \a \log k \tau} + C_4 (-i)e^{-i k \tau - i \a \log k \tau}  . \nonumber
\end{eqnarray}

Perturbations well outside the horizon ($|k \tau| \ll$ 1) are given by
\begin{eqnarray}
\Phi_k & = & -\frac{3}{\a}  \frac{c_1}{ (-\tau)} + \frac{\sqrt{9-8 \a^2}-3}{4 \a}  c_3 (- \tau)^{\frac{1}{2} + \frac{1}{2}\sqrt{9 - 8\a^2}}  + \frac{-\sqrt{9-8 \a^2}-3}{4 \a} c_4 (- \tau)^{\frac{1}{2} - \frac{1}{2}\sqrt{9 - 8\a^2}} \nonumber \\
\Psi_k & = & \frac{c_1}{ (-\tau)} + c_2 (-\tau)^2 + c_3 (- \tau)^{\frac{1}{2} + \frac{1}{2}\sqrt{9 - 8\a^2}}  +c_4(- \tau)^{\frac{1}{2} - \frac{1}{2}\sqrt{9 - 8\a^2}}.  \label{late1}
\end{eqnarray}
The $c_1$ term time-shifts the unperturbed trajectory; this is the adiabatic perturbation.   The $c_2$ term translates $\psi$; shifts in  $\psi$ are conserved outside the horizon, but since $\psi$ itself is exponentially accelerating the relative perturbation rapidly becomes irrelevant.
The $c_3$ and $c_4$ terms are massive modes that don't change the angular momentum ($d J_k = 0$) and soon decay away.

To match the sub-horizon $C_i$s onto the super-horizon $c_i$s requires numerically integrating Eqs.~\ref{eq:xeom1} \& \ref{eq:yeom1} through horizon crossing. Performing this integration reveals curious behavior.  $C_1$ and $C_4$ get mapped to exponentially large (in $\a$) $c_1$, whereas $C_2$ and $C_3$ get mapped to exponentially small $c_1$. The perturbations blamelessly follow their deep-inside-the-horizon forms Eq.~\ref{early1} until $k \tau = - \sqrt{2} \a$, then they abruptly start growing (or shrinking) exponentially in $\tau$, 
and then innocently follow their super-horizon forms Eq.~\ref{late1} after $k \tau =-1$. 

The onset of the rapid growth can be partially understood by diagonalizing the potential term in the second-order action; this gives eigenvalues $m^2_{\pm} = \frac{1}{ \tau^2} \left(  k^2 \tau^2 -2 - \a^2 \pm \a \sqrt{16 + \a^2}
 \right)$ so that $m_{-}^2$ becomes tachyonic for $k^2 \tau^2 \, \lsim \, 2h^2$. However this analysis misses the essential role of the Coriolis $\Phi_k \dot{\Psi}_k$ term, which arises because of the motion of the background field and is relevant throughout. Instead, the rapid growth is best understood in terms of the anti-correlation between angular momentum and angular velocity, as is explained below Eq.~\ref{eq:perturbationsummary}. 
 (Note that the rapid growth occurs during horizon crossing, not after as in \cite{Leach:2000yw}.)

Finally, all that remains is to fix the initial values of the $C_i$s by promoting the field to a quantum operator and insisting that on very short scales we recover the regular Bunch-Davies vacuum. As in slow-roll inflation, insisting that there be no particles present at early times puts $C_1 = C_2 = 0$; Heisenberg's uncertainty principle then gives 
\begin{equation}
C_3 = C_4 = \frac{1}{\sqrt{2k}} .
\end{equation}
Numerically integrating this initial condition gives density fluctuations after inflation that are well-approximated by Eq.~\ref{eq:perturbationsummary}. \\

\noindent Readers who have made it this far may also find it interesting to read \cite{Mizuno:2017idt}, a follow-up study to this paper, which confirmed the results above and explored some aspects in more detail.


\end{document}